\documentclass[conference]{IEEEtran}
\IEEEoverridecommandlockouts
\usepackage{algorithm, algorithmic}
\usepackage{amsmath, amssymb, amsfonts}
\usepackage{array}
\usepackage{cite}
\usepackage{graphicx}
\usepackage{textcomp}
\usepackage{url}
\usepackage{xcolor}
\def\BibTeX{{\rm B\kern-.05em{\sc i\kern-.025em b}\kern-.08em
		T\kern-.1667em\lower.7ex\hbox{E}\kern-.125emX}}

\begin{document}
\title{Protecting User Privacy in Online Settings via Supervised Learning}

\author{
    \IEEEauthorblockN{Alexandru Rusescu, Brooke Lampe, and Weizhi Meng}
    \IEEEauthorblockA{SPTAGE Lab, Department of Applied Mathematics and Computer Science, \\Technical University of Denmark, Denmark}
    }

\maketitle

\begin{abstract}
Companies that have an online presence---in particular, companies that are exclusively digital---often subscribe to this business model: collect data from the user base, then expose the data to advertisement agencies in order to turn a profit. Such companies routinely market a service as ``free," while obfuscating the fact that they tend to ``charge" users in the currency of personal information rather than money. However, online companies also gather user data for more principled purposes, such as improving the user experience and aggregating statistics. The problem is the sale of user data to third parties. In this work, we design an intelligent approach to online privacy protection that leverages supervised learning. By detecting and blocking data collection that might infringe on a user's privacy, we can restore a degree of digital privacy to the user. In our evaluation, we collect a dataset of network requests and measure the performance of several classifiers that adhere to the supervised learning paradigm. The results of our evaluation demonstrate the feasibility and potential of our approach.
\end{abstract}


\section{Introduction}

Over the past few decades, the Internet has become a part of people's day-to-day lives. The activities facilitated by the modern Internet are varied and innumerable: browsing recipes, purchasing products, sharing videos, banking, billing, socializing, and many, many more. In the era of Internet-enabled expedience, users tend to overlook the slow decline of privacy in favor of staggeringly greater convenience. However, the discussion of Internet-related privacy infringement is becoming more relevant in the public eye. Some news agencies and governments have started to point out how these privacy-invading practices can affect people's lives---and even small businesses---without them realizing it~\cite{b1}. International incidents have piqued the interest of news outlets and security specialists to look into this new ``surveillance" business model, sparking comparisons between data collection and "Big Brother" from George Orwell's book ``1984"~\cite{b2,b3}.

Privacy can be invaded by any number of means, and there is no clear distinction between that which is permissible and that which is strictly prohibited by law. It has been reported that a majority of Americans believe their online and offline activities are being tracked and monitored by both companies and the U.S. government with some regularity ~\cite{b7}. Invasions of privacy have become the norm, not the exception. In fact, privacy infringement is such a common condition of modern life that approximately 60\% of U.S. adults say they do not think it is possible to go about daily life without having data collected by various companies or the U.S. government.

That said, some countries have made process and even forced companies to limit some of their practices; e.g., the European Union has implemented---and enforces---the General Data Protection Regulation (GDPR)~\cite{gdpr}. Unfortunately, the GDPR is still not enough to defeat some external threats. For example, a malicious third party may exfiltrate data and documents that colleagues create, access, store, and share across an organization. When a third party gains access to an individual's private information, there is a risk of data loss, reputational damage, and regulatory fines.

\textbf{Contributions.} In the literature, we have seen many potential strategies (e.g., privacy-preserving techniques~\cite{Liu2021_iot,Qin2023}) to ensure data privacy in various environments, but safeguarding a user's privacy online is still an open challenge---especially with the rapid pace of digitization. In this work, we contribute to the defense of online privacy by introducing a tool that can be used to block HTTP requests that would gather users' data. Our main contributions are summarized below:

\begin{itemize}
    \item We develop a tool which leverages supervised machine learning to detect malicious online requests and protect users' privacy. In addition, we detail the API implementation of our tool.
    \item In our evaluation, we collect a dataset of online requests and evaluate the performance of three supervised learning classifiers. The results demonstrate the feasibility and potential of our tool.
\end{itemize}

The remainder of this paper is organized as follows: Section~\ref{sec:2} introduces related work on privacy protection. In Section~\ref{sec:3}, we explain our proposed tool in detail, including the data collection process and the three supervised learning classifiers. In Section~\ref{sec:4}, we describe the API implementation of our tool and discuss experimental results. Section~\ref{sec:5} concludes our work.

\section{Related Work} \label{sec:2}
In this section, we highlight various privacy-enhancing schemes, each of which aims to ensure a user's privacy while he or she is online.

Rodrigo-Gin{\'{e}}ss \emph{et al.}~\cite{Rodri2018} crafted a tool called ``PrivacySearch" that contributes to the development of privacy-enhancing technologies (PETs). This paper addresses the problem of personalized profiles based on Web Search Engines (WSEs). Such personalized profiles are created with the intent of selling information. The authors exploit the query-generalization principle: when a user types a query, the PrivacySearch tool replaces the text of the query with generic terms before submitting the generalized query to the WSE. The tool supports three different privacy levels: low, medium, and high. The three privacy levels refer to the degree of generalization provided by the tool; a higher privacy level will generate a more generic query (compared to the original query) than a lower privacy level.

Reiter \emph{et al.}~\cite{Reiter1998} proposed a system, dubbed ``Crowds," to increase the privacy of web transactions. Crowds leverages the concept of a ``crowd," in which one hides one's actions behind the actions of many, many others. Crowds---i.e., the ``crowd" technique---works as follows:

\begin{enumerate}
    \item A user gains access to the system called Crowds
    \item The user's request to a web server is passed to a random member of the same system in an encrypted form
    \item Upon receiving the request, the random member flips a biased coin to determine if he or she should submit the request or forward it to another randomly selected member (the request data is encrypted until the moment the request is submitted to the server)
    \item The previous step repeats until the request is submitted
    \item The web server's reply traverses the same path but in reverse
\end{enumerate}

Reiter \emph{et al.} provide an excellent strategy to anonymize web transactions, though the Crowds strategy comes at the cost of additional overhead. This system also obfuscates the information that a local eavesdropping might use to learn about the identity of the receiver.

Mozilla offers a solution called ``Facebook Container"~\cite{b4} to set boundaries for Facebook and other Meta websites. Facebook Container is an extension that isolates Meta sites (e.g., Facebook, Instagram, and Messenger) from the remainder of the web to limit where the company can track its users. Meta's ``like" and ``share" contain Facebook trackers, and these buttons can be found in numerous websites, from news to shopping to blogs and more. Mozilla's Facebook Container alerts the user when a tracker is discovered on a non-Meta site by adding an icon to the address bar and subsequently blocking the trackers. Users are given the option to disable Facebook Container on specific websites, allowing Facebook to see their activity there. Mozilla's tool constrains the volume of data that Meta is able to obtain, though other advertisers might still be able to correlate Facebook activity with a user's regular browsing.

Another solution, ``Pi-Hole"~\cite{b5}, enhances digital privacy by blocking known advertising domains. Pi-Hole was conceived as an open-source alternative to Ad-Trap. A Raspberry Pi---a small, single-board computer---acts as a Domain Name System (DNS) server for a given private network. DNS servers are populated with the mappings of domain names to IP addresses. As such, Pi-Hole comes with a file of blacklisted hosts~\cite{b6}, which is properly maintained and up to date. Whenever queries are made to known advertising domains, a falsified IP address is given to the web server. Then, instead of an advertisement, the web server serves a tiny blank image file or web page. A user can surf the web freely and privately, knowing that an invasive request will be misdirected and will never reach its target IP address. Thus, the user protected from sharing private information (e.g., browsing habits) with unwanted sources.

\section{Our Approach} \label{sec:3}
\begin{figure}
\centering
\includegraphics[width=.95\linewidth]{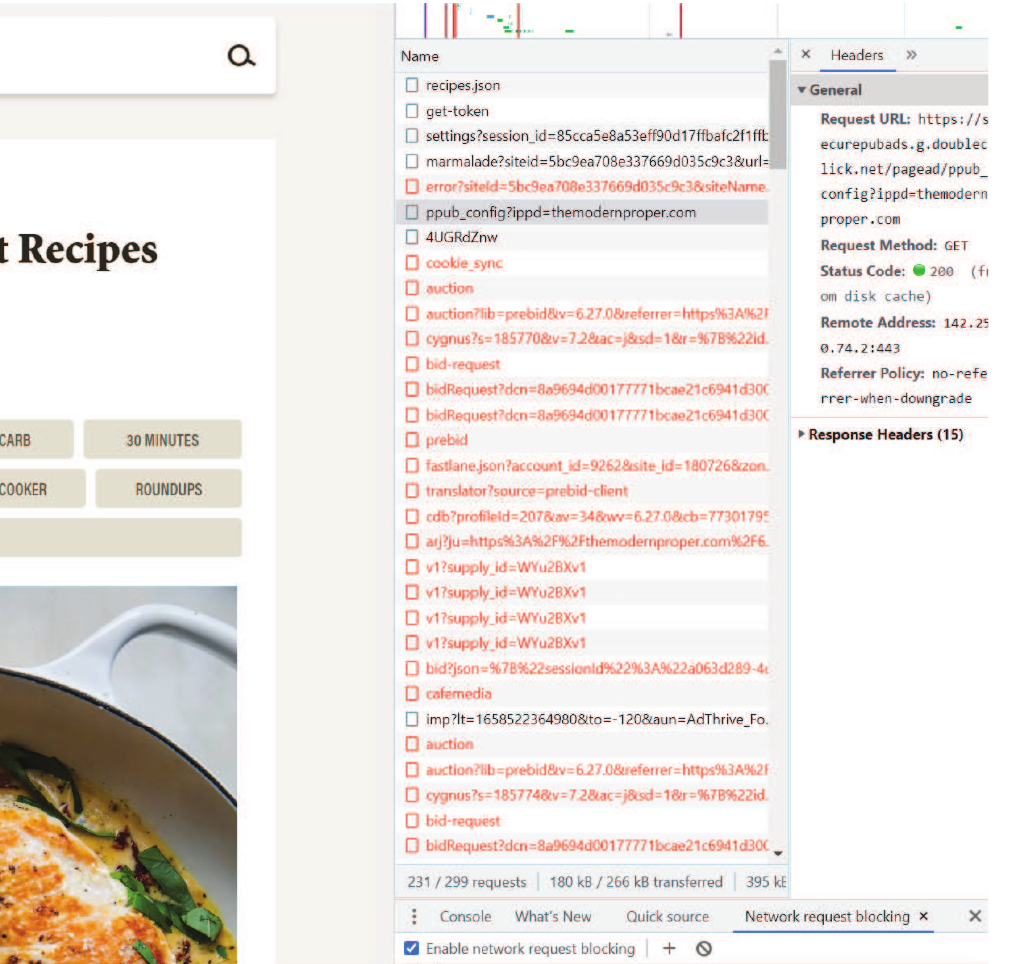}
\caption{A network request log from a cooking website.}
\label{fig:1}
\vspace{-3mm}
\end{figure}

Figure~\ref{fig:1} provides a snapshot of the network requests exchanged by a cooking website. A number of the requests are utterly irrelevant to the usability of the site, while a select few are necessary in order to supply the information that a user seeks. More than 80\% of the network requests are involved in delivering information to third-party advertisers who can do what they please with the data.

Going a step further, we reviewed and analyzed the network requests of multiple websites. We inspected the network logs and verified that suspicious requests were indeed transmitted to advertisement domains or other domains that had no relevance to the website in question.

\begin{figure}
\centering
\includegraphics[width=.95\linewidth]{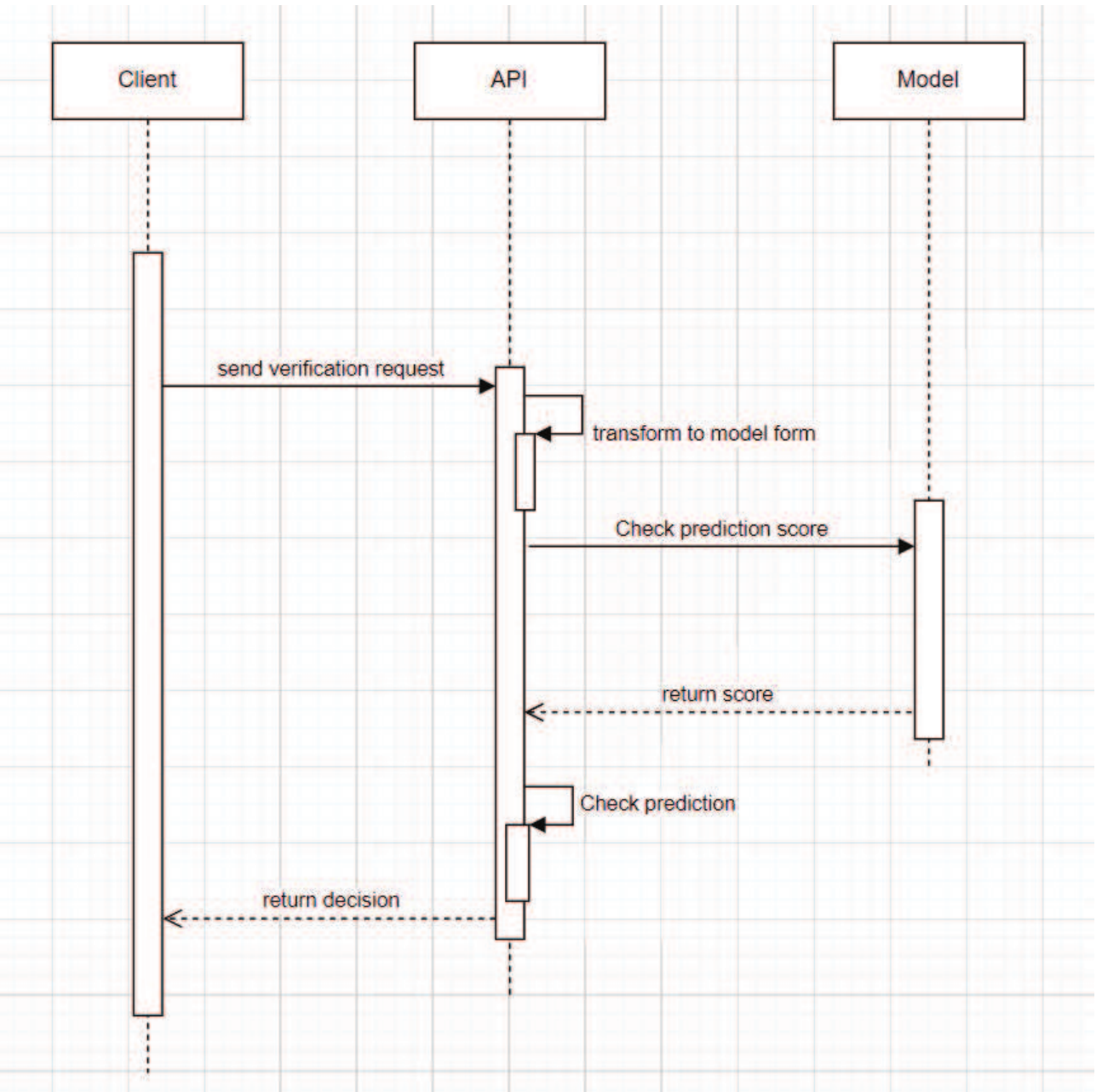}
\caption{A sequence diagram outlining our proposed approach.}
\label{fig:2}
\vspace{-3mm}
\end{figure}
	
\subsection{Design Phase}
A sequence diagram is depicted in Figure~\ref{fig:2}, providing an abstract view of our design. To ensure that our proposed solution would be accessible, we decided to expose an endpoint. This endpoint would be responsible for receiving requests, pre-processing those requests, and invoking the tool. We implemented our endpoint as an application programming interface (API). This API receives an input, runs that input through our tool (a supervised machine learning model), and returns an output.

To build the tool itself, we needed to select a suitable machine learning model. The model would be expected to differentiate between benign traffic and privacy-invading traffic. In this work, we considered three supervised learning classifiers due to their popularity and their reportedly good performance.

\begin{itemize}
    \item \textbf{Logistic Regression (LR).} This is a type of statistical modeling often used for classification and for predictive analytics.~\cite{Kim2009}. Logistic regression falls under the umbrella of linear regression, but it is a special case: in logistic regression, the predicted outcomes are categorical. If there are two possible outcomes, it is referred to as binomial logistic regression, and if there are more than two possible outcomes, then it is called multinomial logistic regression~\cite{Ouyed2020}. During logistic regression, the model first computes the sum of the input features and then applies the logistic function. The output is guaranteed to be between 0 and 1. In our scenario, 0 would be ``benign" and 1 would be ``privacy-invading." The closer the value is to 1, the higher the probability that the current sample will be categorized as privacy-invading, and vice versa.
    \item \textbf{Decision Tree.} This is a supervised learning technique that operates on a tree-like structure: the nodes represent the features of the dataset, the branches represent the decision rules, and the leaf nodes represent the outcomes~\cite{Bahnsen2015,Li2015_icc}. To construct a decision tree, each feature in the dataset is mapped to a node. Starting at the root node, branches with decision rules are created to extend to the next nodes, each of which becomes a subtree. This process continues for each subtree that is created~\cite{Wang2019}.
    \item \textbf{Support Vector Machine.} This supervised learning scheme uses data points plotted in an $n$-dimensional space, where $n$ represents the number of features. Each feature corresponds to the value of a particular coordinate. During the training process, a hyperplane is constructed to subdivide all samples into one of two classes~\cite{Li2019_ispec,Liu2019}. The goals of SVMs can be summarized as follows~\cite{Tavara2019}:
        \begin{enumerate}
            \item Maximize the margin which separates the two classes
            \item Use relatively few training samples---or support vectors---to define the hyperplane which separates the two classes
            \item Classify data that is not linearly separable---with the help of a kernel
        \end{enumerate}
\end{itemize}

All of the classifiers were implemented in Python.

\subsection{Data Collection}
High-quality data is essential to machine learning. The quality of the training data directly influences the model's performance.

As such, during the data collection process, we thoroughly investigated each of the logged network requests to determine which requests were required and which were not. It was crucial to properly validate the ``required" requests (i.e., confirm that they were necessary), just as it was critical to ascertain that a supposed data-gathering request was purely gathering data and was \textit{not} performing some sort of usability-related function. If data-gathering requests were mislabeled as benign requests, the model might produce false negatives; conversely, if benign requests were mislabeled as data-gathering requests, the model might produce false positives. If benign requests are mistakenly blocked, the user's browsing experience will be disrupted; if privacy-invading requests are mistakenly permitted, then the tool will fail to safeguard the user's privacy.

To attenuate such issues, we perform a particular two-stage test during our data gathering process. Given a request, we scrutinize the payload to determine what type of data will be transmitted to the server. Then, we check the domain name of the server: Is this domain name relevant to the website at hand? If the payload appears suspicious \textit{and} the domain name of the server is unrelated to the current website, then the $curl$ (Client URL) request will be copied to a separate backup file and the domain name will be blocked. Finally, we refresh the site in order to ensure that blocking the suspicious requests does not interfere with the usability of the current website.

If a request is deemed non-invasive (in terms of privacy), then the $curl$ request will still be copied order to provide the machine learning model with examples of both benign and privacy-invading requests.

Once we had collected a sufficient volume of data for training and evaluation, we transferred the data to a spreadsheet for easier accessibility when working with the model. During this process, we manually examined each request, including its URL, payload, and properties. The data was organized in the spreadsheet as follows:

\begin{itemize}
    \item Request data that was \textit{not} extracted from payload (e.g., properties)
        \begin{itemize}
            \item \verb+invasive+ - a binary value that indicates whether or not a request was deemed privacy-invasive
            \item \verb+url+ - a string that specifies the domain name of the request's intended destination
            \item \verb+req_type+ - the type of request (e.g., GET, POST, PUT)
            \item \verb+is_json+ - the format of the payload---either JSON or non-JSON (at this time, for easier data collection, we limit our scope to JSON-formatted payloads)
        \end{itemize}
    \item Request data that was extracted from the payload
        \begin{itemize}
            \item \verb+pl_isprebid+
            \item \verb+pl_appid+
            \item \verb+pl_domain+
            \item \verb+pl_imp+
            \item etc.
        \end{itemize}
\end{itemize}

For the initial implementation of our tool, we limit our scope to payloads that are formatted using the $JSON4$ clear text standard.

\subsection{Data Processing}

\textbf{Data cleaning.}
The quality of a machine learning model is contingent on the quality of the data. As such, we applied several cleaning and pre-processing techniques to the data, as described below~\cite{Wang2020}:

\emph{Noise removal.}
``Noise" refers to (1) unwanted, meaningless data and (2) unwanted, meaningless perturbation. As a data pre-processing step, noise removal ensures that the data is free of interference, distortion, or uninformative values; that is, noise removal ensures that the data is clean.

In this step, we checked the dataset for duplicate entries. In addition, since the data was manually collected, we reviewed the dataset for human error.

\emph{Outlier filtering.}
``Outliers" are values which do not fall in the average range defined by existing data points. Outliers in datasets can often be attributed to measurement error during data collection, but they can also occur naturally; some data properties are innately prone to outliers.

Much of our data is one-hot encoded; that is, the variables have been converted to binary indicators. The one-hot encoding process implicitly completes the data filtering step.

\emph{Structural error correction \& missing value correction.}
``Structural error" occurs mainly due to naming and spelling discrepancies. If naming conventions are not consistent across the dataset, then the machine learning model---which receives the data as input---might become muddled and inaccurate. Naming and spelling-related discrepancies can arise from typographical errors or from variations in the naming and spelling conventions used by different researchers. In our case, a sample will either be classified as ``benign" or ``privacy-invading." A simple spelling error could cause a benign sample to be labeled privacy-invading, or vice versa.

Missing values differ from structural error; a ``missing value" occurs when no value is stored in a given attribute of a data object. If, during data collection, a request lacks a response, then one or more missing values might result.

We manually corrected for both structural error and missing values while collecting and populating the dataset. During the implementation phase, we decided to remove the \verb+url+ column from the dataset, as it is a categorical column with very few values that repeat themselves. In addition, we elected to one-hot encode the \verb+req_type+ column; we renamed the column \verb+GET/POST+ and converted the values to binary. Since our current dataset contains GET and POST requests, exclusively, the \verb+req_type+ column (now the \verb+GET/POST+ column) was an ideal candidate for the one-hot encoding tactic, reducing clutter in the dataset.

\textbf{Data splitting.}
When it comes to training and testing data, the convention in the literature is the 70-30 split, meaning that 70\% of the samples are used to train the model, while the remaining 30\% are used to test the model.

Therefore, we separated the training column and the target column (which identifies a sample as ``benign" or ``privacy-invading"), creating four new variables:

\begin{itemize}
    \item \verb+x_train+ - This variable contains 70\% of the samples in the dataset, including all feature columns. These are the training samples.
    \item \verb+x_test+ - This variable contains 30\% of the samples in the dataset, including all feature columns. These are the testing samples.
    \item \verb+y_train+ - This variable contains 70\% of the target values in the dataset, which correspond to \verb+x_train+. These are the training targets.
    \item \verb+y_test+ - This variable contains 30\% of the target values in the dataset, which correspond to \verb+x_test+. These are the testing targets.
\end{itemize}

\section{Implementation and Evaluation} \label{sec:4}
In this section, we detail the API implementation and discuss the evaluation results.

\subsection{API implementation}
An application programming interface (API) is an intermediate piece of software that facilitates communication between two applications. This type of software interface is similar to a contract between two parties: There is an expected request structure, and the response adheres to a predetermined format. An API abstracts the particulars of an application and exposes only necessary services to other applications. As such, an API can easily share an application's functionality with external applications, clients, or services.

In this work, a web API was implemented. A web API is a service that runs on a machine and can be accessed by potential clients via HTTP requests. To use our tool, a potential client would send a data transfer object (DTO)---which carries data between processes---to the web API. The tool's response would contain either a ``1" for invasive requests or a ``0" for non-invasive requests.

To implement the web API, we leveraged the Python programming language. A contributing factor in our choice of programming language was Python's compatibility with other components of our tool. Our web API relies on a Python framework called Flask~\cite{Flask}. Flask provides tools and features that help its users to easily create and operate web applications. A simple web server can be constructed with as little as five lines of code.

When the application executes, it will first load the previous model using the \verb+pickle+ module. The previously built model is then made available in the code. The application exposes two endpoints under the routes \verb+/api/predict/lr+ and \verb+/api/predict/dt+. The former generates a prediction based on Logistic Regression, while the latter leverages a Decision Tree to determine if a given request is invasive (``1") or benign (``0"). Both endpoints expect a request with a specific $JSON$-formatted DTO.

The endpoints are strict; all variables are expected to be present in the payload. If the endpoints were not strict---i.e., if the endpoints generated predictions based on incomplete payloads---then the endpoints would be liable to output incorrect results. The required DTO should not be difficult to construct, as it consists of information that the client can extract from the original request.

\begin{figure}
\centering
\includegraphics[width=1.0\linewidth]{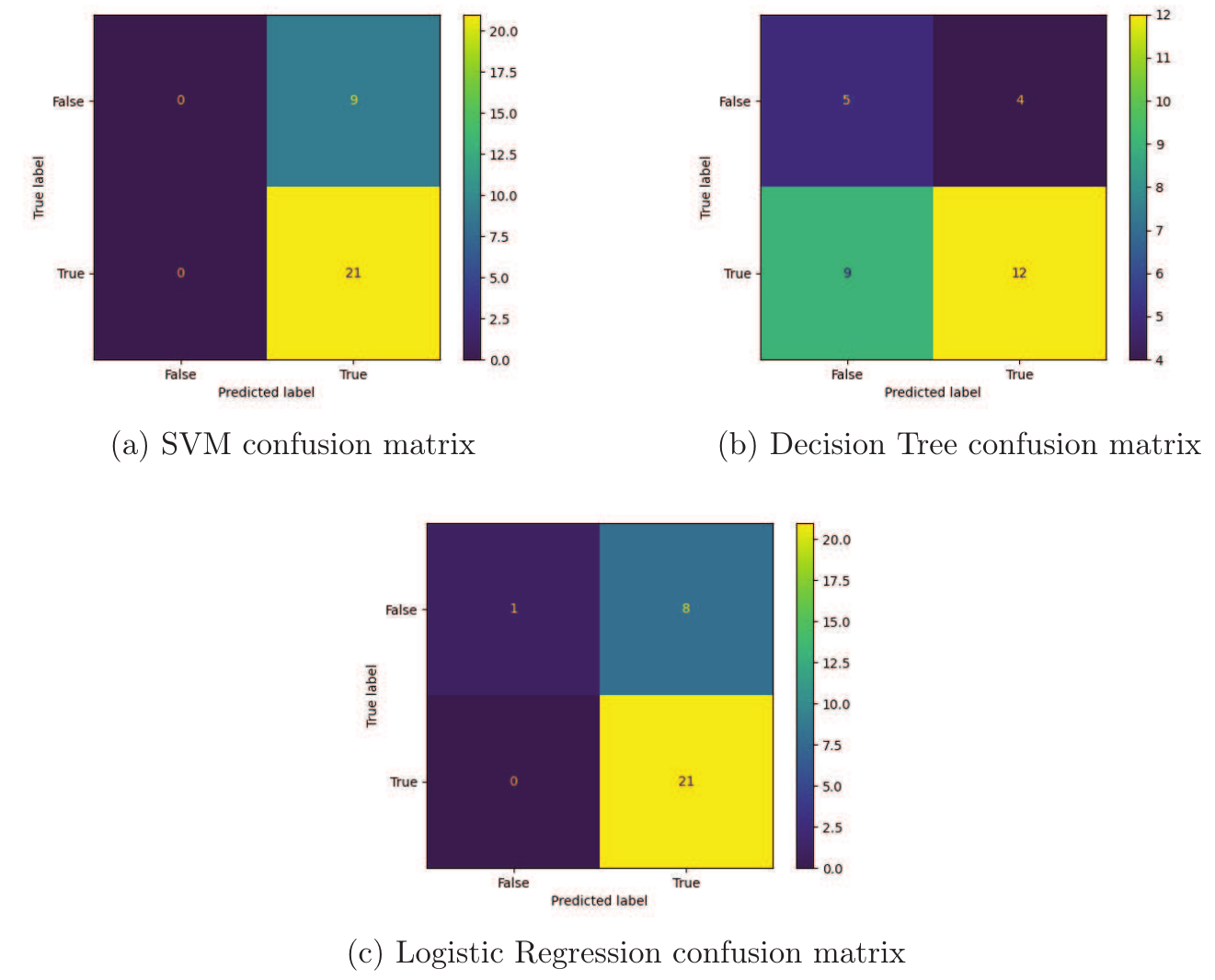}
\caption{Confusion matrices for the support vector machine, decision tree, and logistic regression models.}
\label{fig:eval:network}
\end{figure}

\begin{table*}
\centering
\caption{Evaluation Results} \label{fig:tabl}
\begin{tabular}{p{3cm} p{0.5cm} p{0.5cm} p{0.5cm} p{0.5cm} p{1.25cm} p{1.25cm} p{1.25cm} p{1.25cm} p{1.25cm}}
\hline
    & TP & TN & FP & FN & Accuracy & Precision & Recall & Specificity & F1-score \\ \hline
Support Vector Machine  & 0 & 21 & 9 & 0 &     0.7 &     0.7 &     1.0 &     0.0 & 0.82352 \\ \hline
Decision Tree           & 5 & 12 & 4 & 9 & 0.56666 &    0.75 & 0.57142 & 0.55555 & 0.64864 \\ \hline
Logistic Regression     & 1 & 21 & 8 & 0 & 0.73333 & 0.72413 &     1.0 & 0.11111 &    0.84 \\ \hline
\end{tabular}
\end{table*}

\subsection{Evaluation Results}
To assess the performance of our design, we adopted a confusion matrix---i.e., a table
that provides data regarding the performance of an algorithm. Confusion matrices contain the following four values:

\begin{itemize}
    \item \textbf{True positives (TP).} The number of correctly predicted positive outcomes based on the
    predictive model.
    \item \textbf{True negatives (TN).} The number of correctly predicted negative outcomes based on the
    predictive model.
    \item \textbf{False positives (FP).} The number of incorrectly predicted positive outcomes based on the predictive model.
    \item \textbf{False negatives (FN).} The number of incorrectly predicted negative outcomes based on the predictive model.
\end{itemize}

These metrics are the building blocks for a number of popular performance measures:

\begin{itemize}
    \item \textbf{Accuracy.} Accuracy measures how often the model correctly classifies a sample. It is one of the most commonly used metrics for evaluating models. That said, it is a biased measure and is not necessarily the best indicator of overall performance.
    \item \textbf{Precision.} Precision---or positive predictive value---is defined as the proportion of positive predictions that were actually correct. The formula for precision divides the number of true positives by the total number of positive predictions (true positives and false positives).
    \item \textbf{Recall.} Recall---or sensitivity---is defined as the proportion of actual positives that were predicted correctly. The formula for recall divides the number of true positives by the total number of positive samples in the dataset (true positives and false negatives).
    \item \textbf{Specificity.} Specificity is defined as the proportion of actual negatives that were predicted correctly. The formula for specificity divides the number of true negatives by the total number of negative samples in the dataset (true negatives and false positives). Specificity is similar to sensitivity, except that the perspective shifts from positive to negative.
    \item \textbf{F1-score.} F1-score is calculated as the ``harmonic mean" of precision and recall. Both false positives and false negatives are factored into the F1-score; as such, it is a good metric for imbalanced datasets.
\end{itemize}

The dataset consists of 90 unique requests that were manually collected from various websites, and our supervised machine learning model was trained on 178 unique classes. It is important to note that a relatively small dataset can pose a problem to the predictive ``power" of a machine learning algorithm, since a machine learning model's ability to recognize patterns is generally proportional to the size of the dataset. Smaller datasets correspond to less powerful and less accurate machine learning models~\cite{Kokol2022}.

We evaluated three supervised machine learning paradigms: logistic regression, decision tree, and support vector machine. All three models yielded a result in a very short amount of time. The size of the training dataset was 60 samples, and the size of the testing dataset was 30 samples. Figure~\ref{fig:eval:network} illustrates the confusion matrices that were generated by each of the three models. The results can be interpreted as follows:

\begin{itemize}
    \item The confusion matrix for the support vector machine is depicted in Figure~\ref{fig:eval:network}(a). The model predicted \verb+TRUE+ for all samples, but it was wrong 30\% of the time. The evaluation criteria, as well as the numerical evaluation results, are presented in Table~\ref{fig:tabl}. The model demonstrated a precision of 0.7. The model achieved a perfect 1.0 score for recall, as it correctly identified all positive samples as invasive. Unfortunately, the model achieved this score by classifying all samples as positives, meaning that the model has a serious problem with false positives. These false positives are reflected in the model's specificity score, 0.0, which is the lowest possible score. The model achieved an F1-score of approximately 82\%. We can see that the support vector machine has a severe bias toward privacy-invading predictions, which causes it to routinely misclassify benign samples as invasive.

    \item The confusion matrix for the decision tree is shown in Figure~\ref{fig:eval:network}(b). The model that was trained using the decision tree algorithm gave accurate predictions 56.(6)\% of the time and inaccurate predictions 43.(3)\% of the time. While $\approx56\%$ is better than 50-50 odds, it is not better by much. The precision of the model was 0.75---somewhat higher than the precision of the support vector machine. The recall, however, was 0.57, which is much lower than the 1.0 recall of the support vector machine. When it came to specificity, the model attained a modest 0.55(5), indicating that 55\% of the non-invasive requests would be correctly whitelisted by the model. Note that the decision tree produced the fewest false positives of all the models; as such, it would be least likely to interfere with the usability of a given website. Unfortunately, the model's F1-score was 0.64, which is quite low for a machine learning model. The low F1-score can generally be attributed to the relatively high number of false negatives; the decision tree model misclassified nine invasive samples as benign. Neither the support vector machine nor the logistic regression model produced any false negatives.

    \item The confusion matrix for the logistic regression is illustrated in Figure~\ref{fig:eval:network}(c).
    The logistic regression-based model generated correct predictions for 73.(3)\% of the samples and incorrect predictions for 26.(6)\% of the samples. At a precision of 0.72, the logistic regression model sits between the support vector machine and the decision tree in terms of positive predictive value. The logistic regression model matched the support vector machine in terms of recall, achieving a perfect 1.0 score. Unfortunately, the logistic regression model has the same issue with false positives as the support vector machine---non-invasive requests are regularly misclassified as invasive. Fortunately, the logistic regression model achieved a somewhat better specificity score than the support vector machine---0.1(1) instead of 0.0. The logistic regression model achieved the highest F1-score of all the models: 84\%.
\end{itemize}

\subsection{Limitation and Open Challenges}
As earlier discussed, each of the models---support vector machine, decision tree, logistic regression---comes with both advantages and disadvantages. In terms of recall (proportion of invasive requests that were correctly identified), the worst performing model was the decision tree with a score of 0.57. For the decision tree model, we can expect that approximately 43\% of invasive requests might pass as non-invasive. 43\% might seem quite high at first glance, but the current state of user privacy is much worse: for the average user, 100\% of invasive requests will be allowed. Therefore, if we can block 57\% of invasive requests, then we have made a significant step forward in terms of user privacy. The issue, then, is the false positives. If our tool blocks non-invasive requests that a website needs in order to function properly, then usability will be impacted. The specificity score of the decision tree was 0.55(5). As such, we anticipate that approximately 44\% of non-invasive requests will be blocked, impeding a given website's ability to run.

The next model we need to address is the model trained using the support vector machine algorithm. On the surface, the model's accuracy, precision, and recall are all reasonable---0.7, 0.7, and 1.0, respectively. However, the SVM-based model would be impracticable in a real-world scenario due to its bias toward positive predictions. Under evaluation, the model incorrectly classified all the benign requests as invasive. If a user were surfing the web, the model would block all requests, both invasive and non-invasive, completely disrupting the user's day-to-day activities on the web.

Finally, we will review the logistic regression-based model. Similar to its support vector machine-based counterpart, it is exceedingly biased toward positive predictions. As such, it suffers similar limitations in a real-time scenario; it would disrupt a user's web-surfing experience by blocking most---if not all---non-invasive requests.

\section{Conclusion} \label{sec:5}
The objective of privacy-enhancing technologies (PETs) is to safeguard user data from misappropriation and mishandling. When a user's data is protected, a user's privacy will be protected as well. This project was built upon the following hypothesis: \textit{It is possible to enhance the privacy of normal users without disrupting their day-to-day web surfing activities.} We set out to develop a proof-of-concept tool to confirm our hypothesis. We constructed a supervised machine learning model that classifies HTTP requests as invasive or non-invasive, and we evaluated three different machine learning paradigms as the foundation of the model. Our results demonstrate that our current solution has the potential to be used in real life---but at the cost of browsing convenience and usability. Future work involves collecting a much larger dataset and training new models to further improve the performance metrics.

\section*{Acknowledgment}
This research work was funded by the European Union's H2020 DataVaults project with GA Number 871755.

\end{document}